\documentclass[superscriptaddress,longbibliography,aps,prl,reprint,showpacs]{revtex4-1}

\usepackage[english]{babel}
\usepackage[T1]{fontenc}
\usepackage{graphicx}
\usepackage{amsmath}
\usepackage{soul}
\usepackage{natbib}
\usepackage{eqnarray}
\bibstyle{aapmrev4-1}
\begin{document}
\title{Ultrafast X-ray Diffraction Thermometry Measures the Influence of Spin Excitations on the Heat Transport through nanolayers}
\author{A. Koc} \author{M. Reinhardt}
\affiliation{Helmholtz Zentrum Berlin, Albert-Einstein-Str. 15, 12489
  Berlin, Germany}
\author{A. von Reppert} %\author{J. Pudell}
\affiliation{Institut f\"ur Physik \& Astronomie,
  Universit\"at Potsdam, Karl-Liebknecht-Str. 24-25, 14476 Potsdam,
  Germany}

\author{W. Leitenberger}
\affiliation{Institut f\"ur Physik \& Astronomie,
  Universit\"at Potsdam, Karl-Liebknecht-Str. 24-25, 14476 Potsdam,
  Germany}

\author{K. Dumesnil}
\affiliation{Institut Jean Lamour (UMR CNRS 7198), Universit\'e Lorraine, Boulevard des Aiguillettes B.P. 239, F-54500 Vandoeuvre les Nancy c\'edex,   France}

\author{P. Gaal}
\affiliation{Helmholtz Zentrum Berlin, Albert-Einstein-Str. 15, 12489
  Berlin, Germany}\affiliation{Institut f\"ur Nanostruktur- und Festk\"orper Physik, Univesit\"at Hamburg, Jungiusstr. 11,20355 Hamburg,Germany}
\author{F. Zamponi}
\affiliation{Institut f\"ur Physik \& Astronomie,
  Universit\"at Potsdam, Karl-Liebknecht-Str. 24-25, 14476 Potsdam,
  Germany}
\author{M. Bargheer} \email{bargheer@uni-potsdam.de}
\homepage{http://www.udkm.physik.uni-potsdam.de} \affiliation{Institut
  f\"ur Physik \& Astronomie, Universit\"at Potsdam,
  Karl-Liebknecht-Str. 24-25, 14476 Potsdam, Germany}
\affiliation{Helmholtz Zentrum Berlin, Albert-Einstein-Str. 15, 12489
  Berlin, Germany}

\newcommand{\superscript}[1]{\ensuremath{^{\textrm{#1}}}}
\newcommand{\subscript}[1]{\ensuremath{_{\textrm{#1}}}}

\date{\today}
\begin{abstract}
    We investigate the heat transport through a rare earth multilayer system composed of Yttrium (Y), Dysprosium (Dy) and Niobium (Nb) by ultrafast X-ray diffraction. This is an example of a complex heat flow problem on the nanoscale, where several different quasi-particles carry the heat. The Bragg peak positions of each layer represent layer-specific thermometers that measure the energy flow through the sample after excitation of the Y top-layer with fs-laser pulses. In an experiment-based analytic solution to the nonequilibrium heat transport problem, we derive the individual contributions of the spins and the coupled electron-lattice system to the heat conduction. The full characterization of the spatiotemporal energy flow at different starting temperatures reveals that the spin excitations of antiferromagnetic Dy speed up the heat transport into the Dy layer at low temperatures, whereas the heat transport through this layer and further into the Y and Nb layers underneath is slowed down. The experimental findings are compared to the solution of the heat equation using macroscopic temperature-dependent material parameters without separation of spin- and phonon contributions to the heat. We explain, why the simulated energy density matches our experiment-based derivation of the heat transport, although the simulated thermoelastic strain in this simulation is not even in qualitative agreement.
\end{abstract}
%\pacs{}
\maketitle
Heat transport at the nanoscale has become an important problem of contemporary physics.\cite{cahi2003,cahi2014a,hoog2015} The field is driven largely by the need to improve heat transport characteristics in integrated circuits operating at high clock rates.\cite{bala2002} The design length scales approach the physical limits, where wave fundamental properties of phonon-heat conduction play an important role.\cite{luck2012,xion2016} Research on the functionality of interfaces in nano-electronics is prevalent, and the heat transport characteristics of interfaces depend strongly e.g. on the roughness of the interface, which is often hard to control in the fabrication process \cite{cahi2003,cahi2014a,chen2004a}.
In many insulators and semiconductors the heat capacity is dominated by phonons, whereas electrons only contribute significantly at high temperatures. The heat transport in metals in contrast is dominated by the conduction band electrons and the excitation of phonons typically reduces the heat transport, because they act as scatterers for electrons \cite{ashc2011a}. In some magnetic materials with strong exchange interactions and large magnetic moments the spin-correlations can contribute more than half of the specific heat over large temperature ranges \cite{pech1996a,grif1954a,gers1957a}. One classical example is the rare earth Dysprosium, which we are investigating in this article. Similar to phonon excitations, the magnetic excitations are known to reduce the heat conductivity when it is dominated by the electrons. \cite{boys1986a} On the other hand heat conduction by magnons may dominate in antiferromagnets.\cite{hess2003}
The transport of heat across interfaces in nanostructures with magnetic and nonmagnetic layers is far beyond what can be safely simulated on an ab-initio basis. The additional degree of freedom given by the quasiparticles of the magnetic excitations presents a very complex problem \cite{beau1996a}.
Additional to the basic understanding of heat transport at the nanoscale,\cite{biel2015} fundamental studies of ultrafast magnetism regarding the possibility of all optical magnetic switching \cite{stam2007a,esch2013a,frie2015a,rett2015a} or the role of spin currents \cite{li2016a,choi2014} will profit from a detailed knowledge about transient temperatures and temperature gradients in such systems. Ultrafast x-ray diffraction has only recently become a tool to measure the transient temperatures in multilayers\cite{navi2013a} and to assign contributions from electrons and phonons to thermal transport.\cite{xu2014}

In this paper we present the results of time-resolved ultrafast X-ray diffraction (UXRD) studies on a complex thin film heterostructure with the layering sequence Y/Dy/Y/Nb/Sapphire. We simultaneously measured the relative Bragg peak shifts of all layers as a direct measure of transient strain $\varepsilon (t)$ after optical excitation of the top Y layer. Using the Gr\"uneisen coefficients derived from the thermal expansion, experimentally measured on the same structure, we extract the time dependent energy densities $\rho^Q_{Y},\rho^Q_{Dy},\rho^Q_{Nb}$ in each layer. When only electrons and phonons carry the heat, the transient temperatures $T_{Y,Nb,Dy}(t)\sim \varepsilon_{Y,Nb,Dy}(t)$ can be direcly read from the measured strains $\varepsilon_{Y,Nb,Dy}(t)$. In the antiferromagnetic state of Dy, a large fraction of the energy resides in spin excitations and we show how to separate the phonon- and spin-contributions ($\rho^Q_{S,P}$) via an analytic decomposition of the measured signal. The initial temperature $T_{i}$ is varied from $136$\,K through the N\'eel temperature  $T_{\text{N}}=180$\,K of Dy up to $276$\,K. We find that the additional presence of anti-ferromagnetic  spin excitations in Dy below $T_{\text{N}}$ speeds up the energy flow from the excited Y layer into the Dy layer, where an additional channel for heat dissipation is present. At the same time, the heat transport through Dy is slowed down as the temperature gradient is decreased when the magnetic excitations scatter the electrons which are the main heat transporting quasi-particles.
A full ab-initio simulation of this complex heat transport problem seems impossible, since the interface-resistances, depending on the perfection of the nanostructure and the coupling constants between electrons, phonons and spin-excitations are unknown. Still, heat transport simulations using bulk values for the thermal conductivities \cite{ho1974a,dobr2009a} can be compared with our measured total energy densities, although the contributions of individual quasiparticles are neglected. We find the nonequilibrium of spins and phonons in the observable of the measured strain.\\
The sample shown in Fig. 1a) is grown epitaxially and consists of a 100 nm thick (0001)-oriented Dy layer encapsulated between two $50$\,nm thick Y films with (0001)-orientation in order to prevent oxidation and to stabilize the helical spin order of Dy \cite{dume1995a}. A $100$\,nm thick Nb buffer layer connects this metallic sandwich stucture to an $\mathrm{Al_2 O_3}$ substrate. The thickness values are derived from the Laue-oscillations around individual Bargg peaks. The penetration depth of $32$\,nm for our excitation pulses at $\lambda = 1030$\,nm wavelength was determined by ellipsometry studies, showing that mainly the upper Y layer is excited.\\
Time-resolved X-ray diffraction measurements were performed at the XPP experimental station at the storage ring BESSY II \cite{navi2012a}. Reciprocal space mapping (RSM) is performed by recording the x-ray photons diffracted from the sample at various incidence angles $\omega$ around the Bragg reflections with a two-dimensional hybrid pixel detector (Pilatus 100k, Dectris Inc.) \cite{rein2016a}.
The large extinction length of hard X-ray pulses at $\lambda = 1.38$\,{$\text{\AA}$} allows for simultaneous detection of diffraction signals from all layers of this structure that is opaque to optical light. As an example, Fig. 1b) displays the broad RSM \cite{schi2013a} of the thin films Y, Dy, and Nb as well as the sharp RSM of the Al$_{2}$O$_{3}$ substrate at $T_i=165$\,K. Fig. 1c shows the diffracted intensity of the out-of-plane scattering vector $Q_z$ obtained by integrating the RSM over the in-plane scattering vector components $Q_{x}$ and $Q_{y}$. The signal broadening of the nanolayers in $Q_z$ is due to the limited layer thickness, whereas the mosaic structure of the crystal is mainly observed as a broadening in the in-plane directions.

\begin{figure}
  \centering
  \includegraphics[width = 8.7 cm]{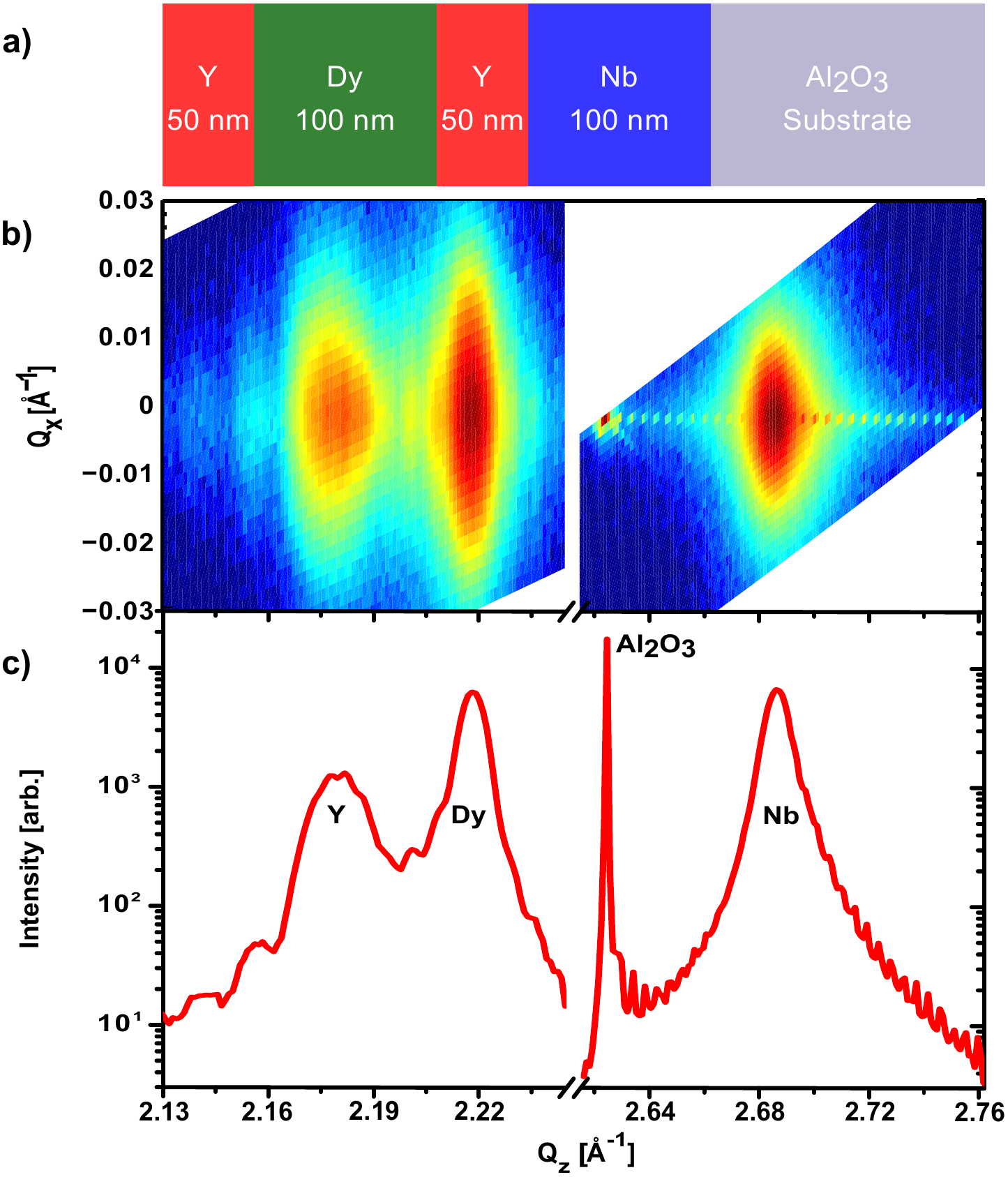}
  \caption{a) Schematic of the sample. b) Reciprocal space map at $T_{i}= 165$\,K of the sample with (0002) reflections of two Y and Dy layers, the (220) reflection of Nb layer and the  (22$\bar 4$0) substrate ($Al_{2}O_{3}$) reflection, respectively. c) Bragg reflections along $Q_z$ obtained by integration of the RSM.
    }
\label{fig:SampleRSM}
\end{figure}

\section{Results}

\subsection{Time-resolved X-ray diffraction data}

\begin{figure}
  \centering
  \includegraphics[width = 8.7 cm, height = 12 cm]{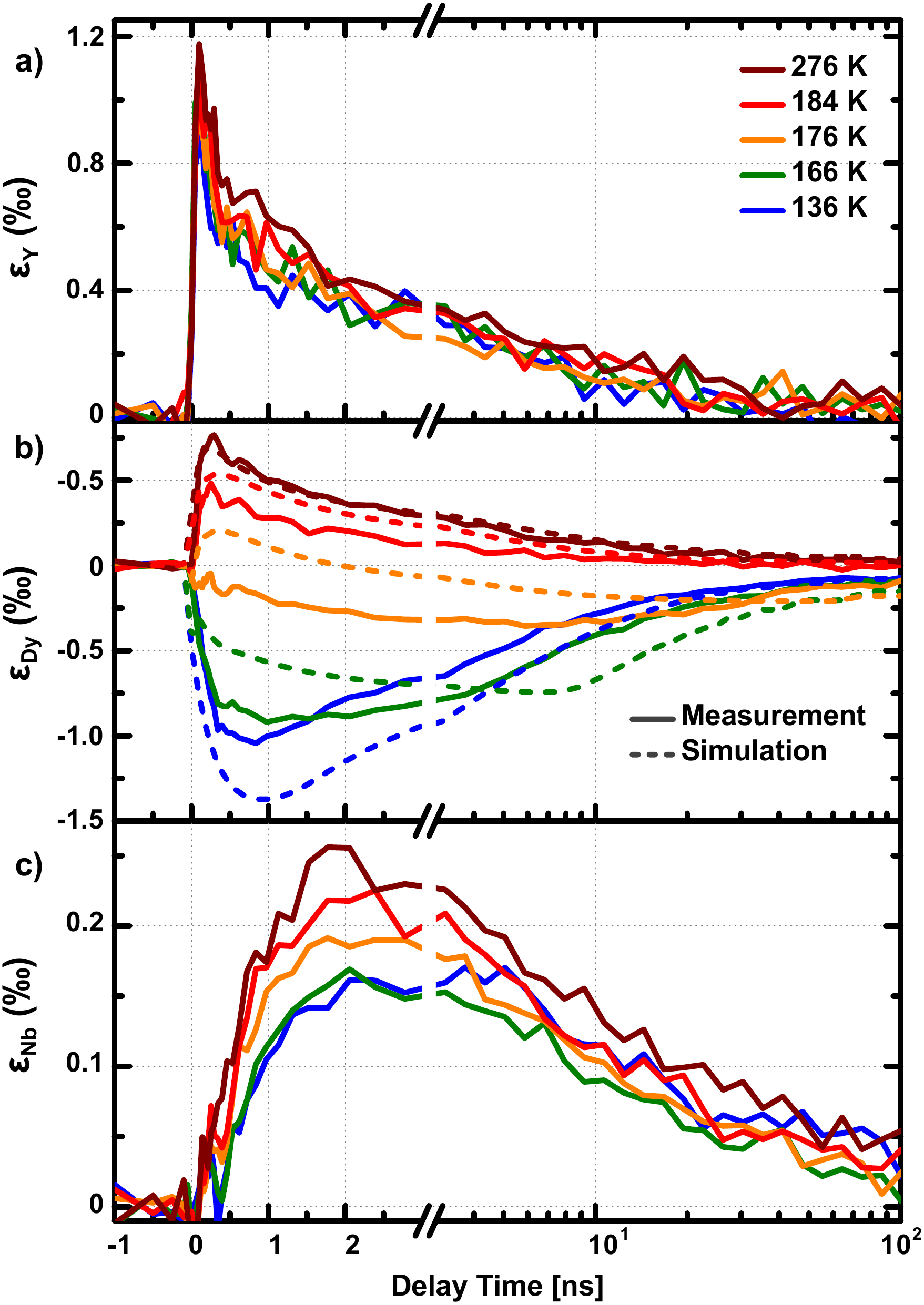}
  \caption{Transient strain $\varepsilon (t)$ for different initial temperatures $T_{i}$ after ultrafast laser heating at $3$\,mJ/$cm^2$ fluence: Solid lines depict the measured transient strain of a) the two Y layers, b) the Dy layer $\varepsilon_{Dy}$ and c) the Nb layer. The dashed lines in panel b) represent the unsuccessful attempt to simulate the strain $\varepsilon_{Dy}^{av}(t)$ by the heat equation, i.e. without taking into account the fact that magnetic excitations and phonons contribute to the heat propagation and the strain.}
\label{fig:UXRDData}
\end{figure}

We present the transient response of each nanolayer after ultrafast laser heating with a laser fluence of 2 mJ/cm$^2$. To extract this information from the data, we fitted each Bragg peak at any delay time with a Gaussian line profile in $Q_z$ to determine the peak position. Transforming the average reciprocal lattice vector $Q_z(t)$ to lattice constants $c(t)=2\pi/Q_{z}(t)$, we obtain the transient strain  $\varepsilon (t) = \frac{c(t)-c(t<0)}{c(t<0)}$ in each layer. In Fig. 2a) the transient strain of both Y layers as the average of the upper and bottom Y layer is shown for different base temperatures $T_i$. At $276$\,K the laser-heated Y layer shows a maximal expansion within the $100$\,ps time-resolution limit given by the pulse duration of the X-rays at beamline.  It relaxes via heat diffusion into the other thin film layers. At lower $T_i$ the same dynamics are observed, however the maximal strain value decreases with decreasing $T_i$. The indirectly heated Dy layer (Fig. 2b) shows very different dynamics depending on the base temperature. At $276$\,K the paramagnetic Dy layer expands and reaches the maximal expansion after about $300$\,ps. In the AFM phase below $T_{\text{N\'eel}}$ the Dy layer contracts upon heating. This negative thermal expansion is a signature of spin-excitations in the antiferromagnetic spin order \cite{repp2016a}. The transient strain in the Nb layer is depicted in Fig. 2c). A maximal expansion of the Nb layer at $T_i=276$\,K is observed at about $1.8$\,ns. At lower base temperatures the maximal expansion shifts to larger time-delays and the magnitude of the maximal expansion is reduced.

\subsection{Data analysis in two-thermal-energies-model}

We analyze the dynamics of the thin film system on timescales larger than the time required for propagating sound through the nanolayer system and smaller than the time for sound-propagation over the in-plane length scale given by the laser-excitation spot. Therefore we assume Hooke's law to be valid, which relates the strain $\varepsilon$ to the stress $\sigma = C_{eff}\,\varepsilon$ via an effective elastic constant $C_{eff}$ \cite{smit1960a,rose1970a,caro1965a} that takes into account the in-plane clamping of the film to the substrate. The macroscopic Gr\"uneisen constant $\Gamma_i=\frac{\alpha_i(T)\,C_{i,eff}}{c_i(T)}$ measures, how efficiently the energy density $\rho^Q_i$ in a subsystem $i$ generates stress $\sigma_i = \Gamma_i \rho^Q_i$. We prefer to write an inverse parameter $\beta_i=\frac{c_i}{\alpha_i}=\frac{C_{eff}}{\Gamma_i}$ which we directly obtained from the bulk specific heat $c_i$ per volume from the literature \cite{jenn1960a,pech1996a,repp2015a} and the expansion coefficient $\alpha_i$ determined from the temperature dependent XRD on the investigated thin film structure. %The linear relation of measured strain to doped energy density is given by the linear thermal expansion coefficient and the spesific heat capacity of the material.
The change of the integral heat
\begin{equation}
\Delta Q_i=V_i\cdot\Delta\rho^Q_i=V_i\beta_i\varepsilon_i
\label{eq:defbeta}
\end{equation}

in a volume $V_i$ of a system is proportional to the lattice strain $\epsilon_i$. At temperatures above $T_{\text{N}}$, the increase of the energy densities $\Delta \rho^Q_{Y,Dy,Nb}$ in Y, Dy and Nb can be directly found from  eq.\ref{eq:defbeta}. Essentially, the energy density of excited phonons in each material drives the lattice expansion, since the electrons carry a negligible fraction of the specific heat, when the electrons have relaxed to approximately the lattice temperature. Table 1 summarizes the $\beta$ constants of Nb, Y, as well as the spin and phonon systems of Dy. These beta values are essentially independent of temperature, as confirmed exemplarily by the constant linear slopes of the curves $\varepsilon_{P,S} \sim Q_{P,S}$ plotted in Fig. 2 of ref. \cite{repp2016a}. To simplify the analysis we do not separately account for the electron and phonon contributions in each metal, since the specific heat of the electron system is always very small. Above $T_N$ the spin contribution to $\beta_{Dy}$ remains constant, but the specific heat of the spins above $T_N$ is very small.

\begin{table}[h!]
\centering
 \begin{tabular}{||c c||}
 \hline
 system & $\beta$ (kJ/cm$^3$) \\ [0.5ex]
 \hline\hline
 Y  & 69  \\
 Dy spin & -20  \\
 Dy phonon & 95  \\
 Nb & 206  \\ [1ex]
 \hline
 \end{tabular}
  \caption{$\beta $ constants of Y,Nb,Dy spin and phonons}
\end{table}

In contrast, the specific heat of the spin system below $T_N$ is very large. In order to measure the individual conributions of phonons and spins to the energy density and expansion of Dy at temperatures below $T_{\text{N}}$, we envoke the tow-thermal-energies model (TTEM) \cite{repp2016a} in Dy. This model assumes that the measured strain $\varepsilon_{\text{Dy}}$ is a superposition of both thermoelastic strain $\varepsilon_{\text{P}}$ and the magnetostrictive strain $\varepsilon_{\text{S}}$.
\begin{align}
\label{eq:EnergyDy}
\Delta \rho^Q_{\text{Dy}} &= \Delta\rho^Q_{\text{S}} +\Delta \rho^Q_{\text{P}} =\beta_{\text{S}}\cdot\varepsilon_{\text{S}}+\beta_{\text{P}}\cdot\varepsilon_{\text{P}}\\
\label{eq:StrainSuperpositionDy}
  \varepsilon_{\text{Dy}} &= \varepsilon_{\text{S}}(\Delta\rho^Q_{\text{S}})
  +\varepsilon_{\text{P}}(\Delta\rho^Q_{\text{P}})
\end{align}
Since the Gr\"uneisenconstants and the coefficients $\beta_{P,S}$ are temperature independent, these strains are a robust and linear measure of the local energy densities.

We can combine eq. \ref{eq:EnergyDy} and \ref{eq:StrainSuperpositionDy} to
\begin{equation}
\Delta\rho^Q_{Dy}=\beta_p(\epsilon_{Dy}-\epsilon_s)+\beta_s\epsilon_s.
\label{deltaQDy}
\end{equation}
Below $T_N$ we have four heat carrying degrees of freedom in the system, namely the spin-excitations in Dy and the phonon-excitation in Dy, Nb and Y. In addition to the three measured transient lattice strains $\varepsilon_{Dy,Y,Nb}$ (Fig. \ref{fig:UXRDData}), we need a fourth equation to find the solution to the heat transport problem.
We conducted temperature dependent ellipsometry measurements proving that the absorbed energy density of the multilayer does not change considerably with temperature. Assuming that no substantial fraction of the initial heat is transported to the substrate, we can identify the total amount of energy deposited in the multilayer at any temperature $\Delta Q_{tot}^{T}$ with the value $\Delta Q_{tot}^{276K}$ measured at $T=276$\,K, where only phonons drive the Dy expansion.
This is an excellent approximation for timescales below $1$\,ns and a very good appoximation up $100$\,ns, because the heat transport into the sapphire substrate is similar for all temperatures.

\begin{figure}
  \centering
 \includegraphics[width = 8.7 cm]{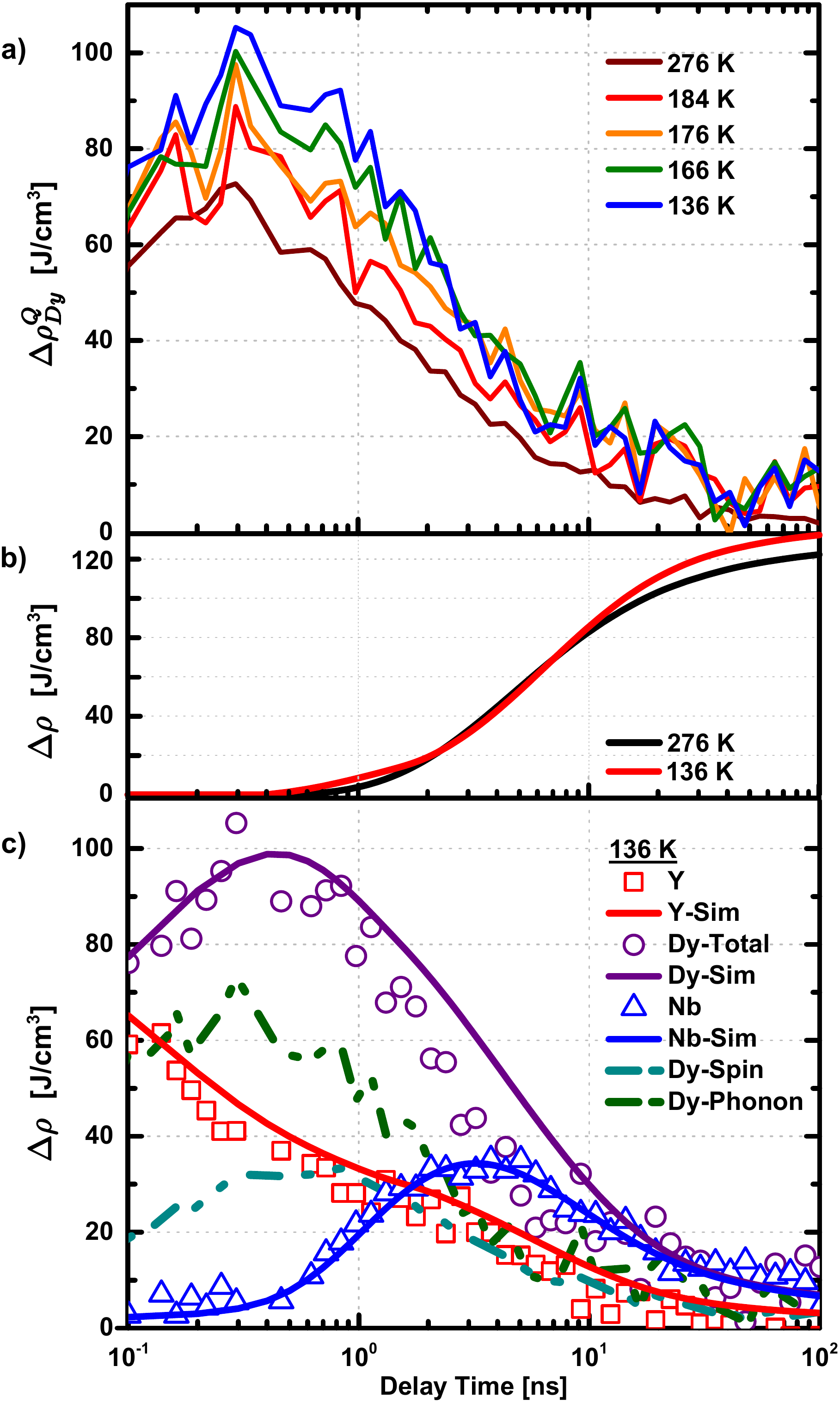}
  \caption{a) Transient increase of the energy density in the Dy layer $\rho^Q_{Dy}$ after optical excitation derived from the measurement according to eq. \ref{deltaQDy}. b) Simulation of the energy density transported into the substrate according to the heat equation.
  c)Symbols show the experimentally determined transient energy densities $\rho^Q_{Y,Dy,Nb}$ in each material. Solid lines represent simulations according to heat equation. Dotted lines show the experimentally derived energy densities in the spin- and phonon system of Dy $\rho^Q_{S,P}$.}
\label{fig:rhoDy}
\end{figure}

When we write
\begin{equation}
\Delta Q_{Dy}^{T}=\Delta Q_{tot}^{276K}-\Delta Q_{Y}^T-\Delta Q_{Nb}^T,
\end{equation}
we only overestimate $\Delta Q_{\text{Dy}}^T$ at low temperatures by the rather small fraction $\Delta Q_{err}=\Delta Q_{tot}^{T}-\Delta Q_{tot}^{276K}$ of heat that is transported into the substrate more than it would be transported at $276$\,K. For convenience, this error can be read from the difference of red and black lines in Fig. \ref{fig:rhoDy}b). The energy densities  $\Delta\rho^Q_{Dy} = \Delta Q_{Dy}^T/V_{Dy}$ in Dy derived for several different base temperatures are plotted in Fig. \ref{fig:rhoDy}a). We find that with lower base temperature a larger and larger fraction of the energy is rapidly transferred from the excited Y layer into Dy.

We now solve eq. \ref{deltaQDy} to obtain equations for the contractive strain $\varepsilon_S$ driven by spin order and the phonon driven expansive strain $\varepsilon_P$, which only depend on measured quantities:

\begin{align}
\label{eq:EpsM}
\varepsilon_{\text{S}} &= \frac{(\Delta\rho^Q_{\text{Dy}} -\beta_{P}\varepsilon_{\text{Dy}})}{(\beta_{S}-\beta_{P})}\\
\label{eq:EpsP}
\varepsilon_{\text{P}}&= \varepsilon_{\text{Dy}}-\varepsilon_{S}
\end{align}

Here, $\Delta\rho^Q_{Dy}$ is the experimentally determined energy density plotted in Fig. \ref{fig:rhoDy}a).
We can now use eq. \ref{eq:defbeta} to derive the contributions to the time-dependent energy densities $\Delta \rho^Q_{S,P}$ in Dy. The corresponding energy densities $\Delta \rho^Q_{Y,Nb}$ of the adjacent layers are determined directly from the measured quantities $\varepsilon_{Y,Nb}$. The resulting energy densities in each material derived from the experiment are plotted in Fig. \ref{fig:rhoDy}c) and compared to a simple calculation of the heat transport according to the heat equation. \cite{schi2014c} Assuming a small thermal interface resistance of 200 MW/m$^2$K only between Nb and Sapphire, we find a very good simultaneous agreement of the experimentally derived total energy density in Dy $\Delta \rho^Q_{Dy}=\Delta \rho^Q_{P}+\Delta \rho^Q_{S}$ and the simulations at $276$\,K and $136$\,K. In contrast, the simulated thermal expansion $\varepsilon_{Dy}^{av}(t)$ averaged over the film thickness (dashed lines in Fig. \ref{fig:UXRDData}b) considerably deviate from the measured strain $\varepsilon_{Dy}$, because the spin- and phonon-system are not even locally in thermal equilibrium. Closer to the phase transition the deviations get stronger.

\section{Discussion}

\begin{figure}
  \centering
  \includegraphics[width = 8.7 cm, height = 12 cm]{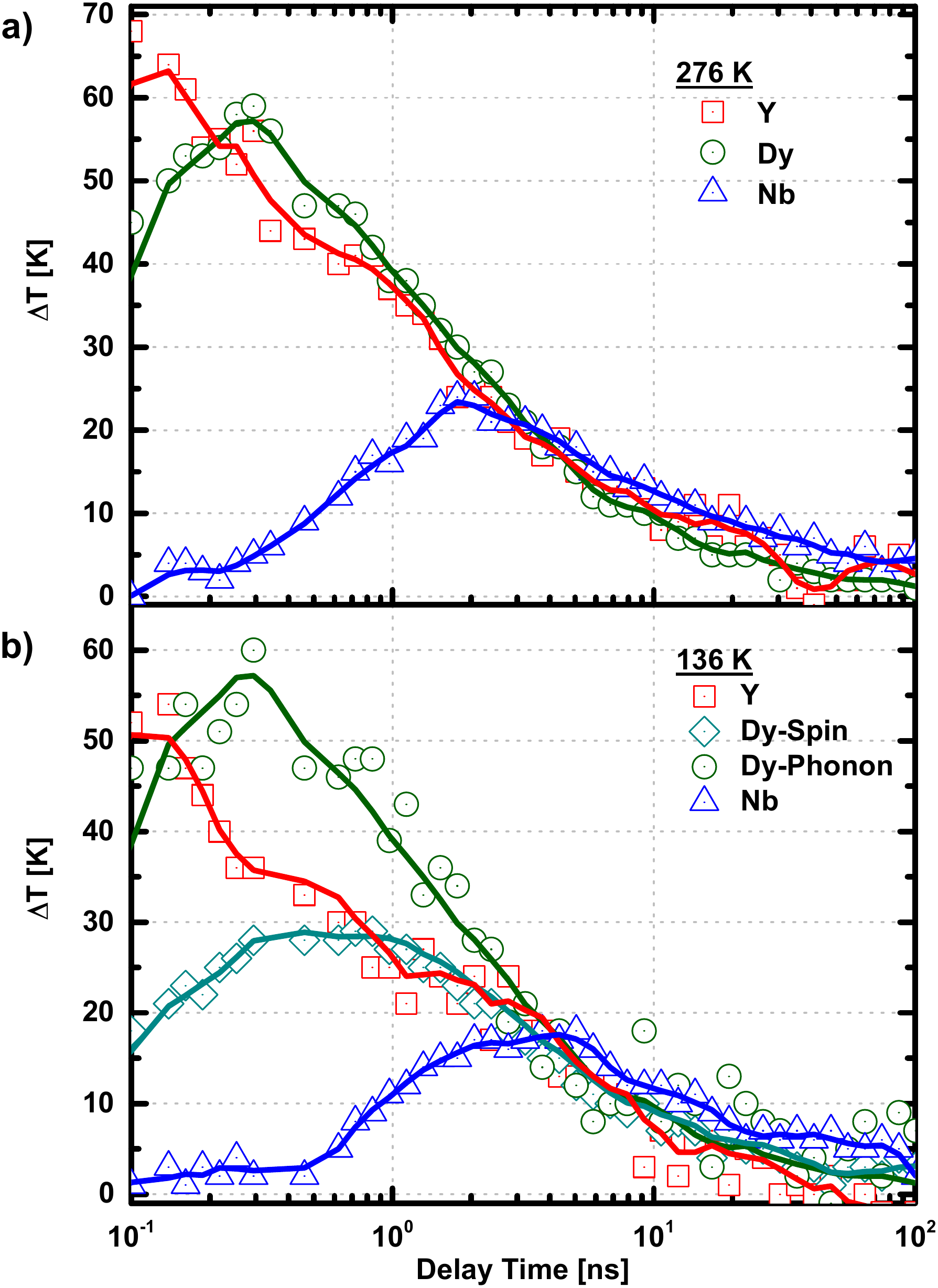}
  \caption{The temperature change in each layer after excitation at a) $276$\,K b) $136$\,K. The temperature in the spin system is only well defined at $T<T_N$. Solid lines represent running averages as a guide to the eye.}
\label{fig:DelT}
\end{figure}

Heat transport is driven by temperature gradients. We therefore plot the transient temperature changes of the spins and phonons $\Delta T_{S,P}$ in Dy and $\Delta T_{Y,Nb}$ in Fig. \ref{fig:DelT} as the experimental solution of the heat transport problem through the three layers as a function of time for two temperatures above and below $T_N$. We use the specific heats $c_i$ of the individual subsystems to calculate the temperature rise $\Delta T$ from $\Delta \rho_Q=\int_{T_0}^{T_0+\Delta T} c(T)dT$. The most striking result is, that within the time resolution of $100$\,ps, we measure that the Y layer is heated by $\Delta T=50$\,K at low and 68\,K at high temperature, although ellipsometry proves that the same amount of energy was deposited by the light pulse. This suggests that the additional energy dissipation channel into spin excitations at low temperatures dramatically speeds up the heat transport across the Y/Dy interface.

Another robust feature seen in Fig. \ref{fig:DelT} is the delay of the temperature rise in the Nb layer, indicating a reduced heat transport through Dy. The temperature rise in the phonon system of Dy at both base temperatures is nearly the same, and therefore  the heat arriving in the spin system effectively is additional to the phonon heat, explaining the observation in Fig. \ref{fig:rhoDy}a) that the increase of the energy density in Dy is higher at low temperature. Note that the kinetics of the temperature rise in the spin- and phonon-systems of Dy are clearly different.

The fact that energy density in the spin system of Dy drives a lattice contraction counteracting the expansion initiated by phonon-heating explains the strong deviations of the observed lattice strain $\varepsilon_{Dy}^{av}(t)\neq \varepsilon_{Dy}$ from the simulated strain (Fig. \ref{fig:UXRDData}b) when spins and phonons are not in a thermal equilibrium. The good agreement on the level of comparing the heat transport can be understood, when we identify the electrons as the main heat transporting quasiparticles. This means that a temperature-gradient in the electron system promotes the transport. Immediately after the optical excitation, the energy is essentially stored in the electron system, with a very large temperature gradient according to the small heat capacity of electrons in Dy. Within the timescale of electron-phonon and electron-spin coupling the heat transport should therefore considerably speed up. This would then lead to an even better match of the simulations with the data in Fig. \ref{fig:rhoDy}a). When the electron-, spin- and phonon temperatures have approached each other, the heat is essentially stored in spin excitations and phonons. Nonetheless it is the electron system which transports the energy, explaining why we simulate the heat transport essentially correct, even if the spin and phonon system have not equilibrated.

\section{Conclusion}

We have exemplified an experimental procedure to measure the heat transport through multilayer systems with thicknesses in the nanometer-range in the non-trivial case, where a considerable fraction of the heat of one material (Dy) is dynamically stored in a strongly interacting spin system. At all temperatures, the heat transport is dominated by electron transport, although electrons only contribute negligibly to the heat capacity. Below the N\'eel temperature, the spin-system opens up an additional heat sink. While the heat transport into the phonon system is nearly unchanged, the spins extract additional energy from the adjacent laser heated Y layer and speed up the inital cooling. At the same time the spin-excitations slow down the electronic heat transport by electron-magnon scattering, which is evidenced by a delayed rise of the Nb temperature, through which the heat is finally dissipated.

Although the average heat, experimentally measured in each layer, is in rather good agreement with standard simulations using the heat equation, there are strong deviations of simulated strain from the measured values. This is because a large fraction of the energy is stored in spin excitations which promote the contraction of the film.
In general, for multilayer-systems where only electrons and phonons carry the heat, the transient temperatures can be directly read from the measured strains, which may be very useful when simulations fail to predict real situations e.g. rough interfaces. For even more complex situations, where several quasiparticles contribute to the heat transport and thermoelastic strain, we have shown how to analytically decompose the measured signal and get the correct decomposition of spin- and phonon contributions to the strain and the heat.
A direct experimental assessment of the heat flow is crucial for understanding the heat transport via various quasi-particles and across interfaces. We believe that such direct experimental crosschecks of theoretical predictions yield valuable information when it comes to optimizing heat transport for real applications.
\section{Acknowledgement}

%\bibliography{BibTeXPRX}

\begin{thebibliography}{36}%
\makeatletter
\providecommand \@ifxundefined [1]{%
 \@ifx{#1\undefined}
}%
\providecommand \@ifnum [1]{%
 \ifnum #1\expandafter \@firstoftwo
 \else \expandafter \@secondoftwo
 \fi
}%
\providecommand \@ifx [1]{%
 \ifx #1\expandafter \@firstoftwo
 \else \expandafter \@secondoftwo
 \fi
}%
\providecommand \natexlab [1]{#1}%
\providecommand \enquote  [1]{``#1''}%
\providecommand \bibnamefont  [1]{#1}%
\providecommand \bibfnamefont [1]{#1}%
\providecommand \citenamefont [1]{#1}%
\providecommand \href@noop [0]{\@secondoftwo}%
\providecommand \href [0]{\begingroup \@sanitize@url \@href}%
\providecommand \@href[1]{\@@startlink{#1}\@@href}%
\providecommand \@@href[1]{\endgroup#1\@@endlink}%
\providecommand \@sanitize@url [0]{\catcode `\\12\catcode `\$12\catcode
  `\&12\catcode `\#12\catcode `\^12\catcode `\_12\catcode `\%12\relax}%
\providecommand \@@startlink[1]{}%
\providecommand \@@endlink[0]{}%
\providecommand \url  [0]{\begingroup\@sanitize@url \@url }%
\providecommand \@url [1]{\endgroup\@href {#1}{\urlprefix }}%
\providecommand \urlprefix  [0]{URL }%
\providecommand \Eprint [0]{\href }%
\providecommand \doibase [0]{http://dx.doi.org/}%
\providecommand \selectlanguage [0]{\@gobble}%
\providecommand \bibinfo  [0]{\@secondoftwo}%
\providecommand \bibfield  [0]{\@secondoftwo}%
\providecommand \translation [1]{[#1]}%
\providecommand \BibitemOpen [0]{}%
\providecommand \bibitemStop [0]{}%
\providecommand \bibitemNoStop [0]{.\EOS\space}%
\providecommand \EOS [0]{\spacefactor3000\relax}%
\providecommand \BibitemShut  [1]{\csname bibitem#1\endcsname}%
\let\auto@bib@innerbib\@empty
%</preamble>
\bibitem [{\citenamefont {Cahill}\ \emph {et~al.}(2003)\citenamefont {Cahill},
  \citenamefont {Ford}, \citenamefont {Goodson}, \citenamefont {Mahan},
  \citenamefont {Majumdar}, \citenamefont {Maris}, \citenamefont {Merlin},\
  and\ \citenamefont {Phillpot}}]{cahi2003}%
  \BibitemOpen
  \bibfield  {author} {\bibinfo {author} {\bibfnamefont {David~G.}\
  \bibnamefont {Cahill}}, \bibinfo {author} {\bibfnamefont {Wayne~K.}\
  \bibnamefont {Ford}}, \bibinfo {author} {\bibfnamefont {Kenneth~E.}\
  \bibnamefont {Goodson}}, \bibinfo {author} {\bibfnamefont {Gerald~D.}\
  \bibnamefont {Mahan}}, \bibinfo {author} {\bibfnamefont {Arun}\ \bibnamefont
  {Majumdar}}, \bibinfo {author} {\bibfnamefont {Humphrey~J.}\ \bibnamefont
  {Maris}}, \bibinfo {author} {\bibfnamefont {Roberto}\ \bibnamefont {Merlin}},
  \ and\ \bibinfo {author} {\bibfnamefont {Simon~R.}\ \bibnamefont
  {Phillpot}},\ }\bibfield  {title} {\enquote {\bibinfo {title} {Nanoscale
  thermal transport},}\ }\href {\doibase 10.1063/1.1524305} {\bibfield
  {journal} {\bibinfo  {journal} {Journal of Applied Physics}\ }\textbf
  {\bibinfo {volume} {93}},\ \bibinfo {pages} {793--818} (\bibinfo {year}
  {2003})},\ \Eprint {http://arxiv.org/abs/http://dx.doi.org/10.1063/1.1524305}
  {http://dx.doi.org/10.1063/1.1524305} \BibitemShut {NoStop}%
\bibitem [{\citenamefont {Cahill}\ \emph {et~al.}(2014)\citenamefont {Cahill},
  \citenamefont {Braun}, \citenamefont {Chen}, \citenamefont {Clarke},
  \citenamefont {Fan}, \citenamefont {Goodson}, \citenamefont {Keblinski},
  \citenamefont {King}, \citenamefont {Mahan}, \citenamefont {Majumdar} \emph
  {et~al.}}]{cahi2014a}%
  \BibitemOpen
  \bibfield  {author} {\bibinfo {author} {\bibfnamefont {D.~G.}\ \bibnamefont
  {Cahill}}, \bibinfo {author} {\bibfnamefont {P.~V.}\ \bibnamefont {Braun}},
  \bibinfo {author} {\bibfnamefont {G.}~\bibnamefont {Chen}}, \bibinfo {author}
  {\bibfnamefont {D.~R.}\ \bibnamefont {Clarke}}, \bibinfo {author}
  {\bibfnamefont {S.}~\bibnamefont {Fan}}, \bibinfo {author} {\bibfnamefont
  {K.~E.}\ \bibnamefont {Goodson}}, \bibinfo {author} {\bibfnamefont
  {P.}~\bibnamefont {Keblinski}}, \bibinfo {author} {\bibfnamefont {W.~P.}\
  \bibnamefont {King}}, \bibinfo {author} {\bibfnamefont {G.~D.}\ \bibnamefont
  {Mahan}}, \bibinfo {author} {\bibfnamefont {A.}~\bibnamefont {Majumdar}},
  \emph {et~al.},\ }\bibfield  {title} {\enquote {\bibinfo {title} {Nanoscale
  thermal transport. ii. 2003-2012},}\ }\href@noop {} {\bibfield  {journal}
  {\bibinfo  {journal} {Applied Physics Reviews}\ }\textbf {\bibinfo {volume}
  {1}},\ \bibinfo {pages} {011305} (\bibinfo {year} {2014})}\BibitemShut
  {NoStop}%
\bibitem [{\citenamefont {Hoogeboom-Pot}\ \emph {et~al.}(2015)\citenamefont
  {Hoogeboom-Pot}, \citenamefont {Hernandez-Charpak}, \citenamefont {Gu},
  \citenamefont {Frazer}, \citenamefont {Anderson}, \citenamefont {Chao},
  \citenamefont {Falcone}, \citenamefont {Yang}, \citenamefont {Murnane},
  \citenamefont {Kapteyn},\ and\ \citenamefont {Nardi}}]{hoog2015}%
  \BibitemOpen
  \bibfield  {author} {\bibinfo {author} {\bibfnamefont {Kathleen~M.}\
  \bibnamefont {Hoogeboom-Pot}}, \bibinfo {author} {\bibfnamefont {Jorge~N.}\
  \bibnamefont {Hernandez-Charpak}}, \bibinfo {author} {\bibfnamefont
  {Xiaokun}\ \bibnamefont {Gu}}, \bibinfo {author} {\bibfnamefont {Travis~D.}\
  \bibnamefont {Frazer}}, \bibinfo {author} {\bibfnamefont {Erik~H.}\
  \bibnamefont {Anderson}}, \bibinfo {author} {\bibfnamefont {Weilun}\
  \bibnamefont {Chao}}, \bibinfo {author} {\bibfnamefont {Roger~W.}\
  \bibnamefont {Falcone}}, \bibinfo {author} {\bibfnamefont {Ronggui}\
  \bibnamefont {Yang}}, \bibinfo {author} {\bibfnamefont {Margaret~M.}\
  \bibnamefont {Murnane}}, \bibinfo {author} {\bibfnamefont {Henry~C.}\
  \bibnamefont {Kapteyn}}, \ and\ \bibinfo {author} {\bibfnamefont {Damiano}\
  \bibnamefont {Nardi}},\ }\bibfield  {title} {\enquote {\bibinfo {title} {A
  new regime of nanoscale thermal transport: Collective diffusion increases
  dissipation efficiency},}\ }\href {\doibase 10.1073/pnas.1503449112}
  {\bibfield  {journal} {\bibinfo  {journal} {Proceedings of the National
  Academy of Sciences}\ }\textbf {\bibinfo {volume} {112}},\ \bibinfo {pages}
  {4846--4851} (\bibinfo {year} {2015})},\ \Eprint
  {http://arxiv.org/abs/http://www.pnas.org/content/112/16/4846.full.pdf}
  {http://www.pnas.org/content/112/16/4846.full.pdf} \BibitemShut {NoStop}%
\bibitem [{\citenamefont {Balandin}(2002)}]{bala2002}%
  \BibitemOpen
  \bibfield  {author} {\bibinfo {author} {\bibfnamefont {A.~A.}\ \bibnamefont
  {Balandin}},\ }\bibfield  {title} {\enquote {\bibinfo {title} {Nanoscale
  thermal management},}\ }\href {\doibase 10.1109/45.985321} {\bibfield
  {journal} {\bibinfo  {journal} {IEEE Potentials}\ }\textbf {\bibinfo {volume}
  {21}},\ \bibinfo {pages} {11--15} (\bibinfo {year} {2002})}\BibitemShut
  {NoStop}%
\bibitem [{\citenamefont {Luckyanova}\ \emph {et~al.}(2012)\citenamefont
  {Luckyanova}, \citenamefont {Garg}, \citenamefont {Esfarjani}, \citenamefont
  {Jandl}, \citenamefont {Bulsara}, \citenamefont {Schmidt}, \citenamefont
  {Minnich}, \citenamefont {Chen}, \citenamefont {Dresselhaus}, \citenamefont
  {Ren}, \citenamefont {Fitzgerald},\ and\ \citenamefont {Chen}}]{luck2012}%
  \BibitemOpen
  \bibfield  {author} {\bibinfo {author} {\bibfnamefont {Maria~N.}\
  \bibnamefont {Luckyanova}}, \bibinfo {author} {\bibfnamefont {Jivtesh}\
  \bibnamefont {Garg}}, \bibinfo {author} {\bibfnamefont {Keivan}\ \bibnamefont
  {Esfarjani}}, \bibinfo {author} {\bibfnamefont {Adam}\ \bibnamefont {Jandl}},
  \bibinfo {author} {\bibfnamefont {Mayank~T.}\ \bibnamefont {Bulsara}},
  \bibinfo {author} {\bibfnamefont {Aaron~J.}\ \bibnamefont {Schmidt}},
  \bibinfo {author} {\bibfnamefont {Austin~J.}\ \bibnamefont {Minnich}},
  \bibinfo {author} {\bibfnamefont {Shuo}\ \bibnamefont {Chen}}, \bibinfo
  {author} {\bibfnamefont {Mildred~S.}\ \bibnamefont {Dresselhaus}}, \bibinfo
  {author} {\bibfnamefont {Zhifeng}\ \bibnamefont {Ren}}, \bibinfo {author}
  {\bibfnamefont {Eugene~A.}\ \bibnamefont {Fitzgerald}}, \ and\ \bibinfo
  {author} {\bibfnamefont {Gang}\ \bibnamefont {Chen}},\ }\bibfield  {title}
  {\enquote {\bibinfo {title} {Coherent phonon heat conduction in
  superlattices},}\ }\href {\doibase 10.1126/science.1225549} {\bibfield
  {journal} {\bibinfo  {journal} {Science}\ }\textbf {\bibinfo {volume}
  {338}},\ \bibinfo {pages} {936--939} (\bibinfo {year} {2012})},\ \Eprint
  {http://arxiv.org/abs/http://science.sciencemag.org/content/338/6109/936.full.pdf}
  {http://science.sciencemag.org/content/338/6109/936.full.pdf} \BibitemShut
  {NoStop}%
\bibitem [{\citenamefont {Xiong}\ \emph {et~al.}(2016)\citenamefont {Xiong},
  \citenamefont {S\"a\"askilahti}, \citenamefont {Kosevich}, \citenamefont
  {Han}, \citenamefont {Donadio},\ and\ \citenamefont {Volz}}]{xion2016}%
  \BibitemOpen
  \bibfield  {author} {\bibinfo {author} {\bibfnamefont {Shiyun}\ \bibnamefont
  {Xiong}}, \bibinfo {author} {\bibfnamefont {Kimmo}\ \bibnamefont
  {S\"a\"askilahti}}, \bibinfo {author} {\bibfnamefont {Yuriy~A.}\ \bibnamefont
  {Kosevich}}, \bibinfo {author} {\bibfnamefont {Haoxue}\ \bibnamefont {Han}},
  \bibinfo {author} {\bibfnamefont {Davide}\ \bibnamefont {Donadio}}, \ and\
  \bibinfo {author} {\bibfnamefont {Sebastian}\ \bibnamefont {Volz}},\
  }\bibfield  {title} {\enquote {\bibinfo {title} {Blocking phonon transport by
  structural resonances in alloy-based nanophononic metamaterials leads to
  ultralow thermal conductivity},}\ }\href {\doibase
  10.1103/PhysRevLett.117.025503} {\bibfield  {journal} {\bibinfo  {journal}
  {Phys. Rev. Lett.}\ }\textbf {\bibinfo {volume} {117}},\ \bibinfo {pages}
  {025503} (\bibinfo {year} {2016})}\BibitemShut {NoStop}%
\bibitem [{\citenamefont {Chen}\ \emph {et~al.}(2004)\citenamefont {Chen},
  \citenamefont {Borca-Tasciuc},\ and\ \citenamefont {Yang}}]{chen2004a}%
  \BibitemOpen
  \bibfield  {author} {\bibinfo {author} {\bibfnamefont {G.}~\bibnamefont
  {Chen}}, \bibinfo {author} {\bibfnamefont {D.}~\bibnamefont {Borca-Tasciuc}},
  \ and\ \bibinfo {author} {\bibfnamefont {R.~G.}\ \bibnamefont {Yang}},\
  }\bibfield  {title} {\enquote {\bibinfo {title} {{Nanoscale Heat
  Transfer}},}\ }\bibfield  {booktitle} {\emph {\bibinfo {booktitle}
  {Encyclopedia of Nanoscience and Nanotechnology}},\ }\href {\doibase
  10.1051/jp4:2005125116} {\ \textbf {\bibinfo {volume} {7}},\ \bibinfo {pages}
  {429--459} (\bibinfo {year} {2004})}\BibitemShut {NoStop}%
\bibitem [{\citenamefont {Ashcroft}\ and\ \citenamefont
  {Mermin}(1976)}]{ashc2011a}%
  \BibitemOpen
  \bibfield  {author} {\bibinfo {author} {\bibfnamefont {N.W.}\ \bibnamefont
  {Ashcroft}}\ and\ \bibinfo {author} {\bibfnamefont {N.D.}\ \bibnamefont
  {Mermin}},\ }\href@noop {} {\emph {\bibinfo {title} {Solid State Physics}}}\
  (\bibinfo  {publisher} {Saunders College},\ \bibinfo {address}
  {Philadelphia},\ \bibinfo {year} {1976})\BibitemShut {NoStop}%
\bibitem [{\citenamefont {Pecharsky}\ \emph {et~al.}(1996)\citenamefont
  {Pecharsky}, \citenamefont {Jr},\ and\ \citenamefont {Fort}}]{pech1996a}%
  \BibitemOpen
  \bibfield  {author} {\bibinfo {author} {\bibfnamefont {V.~K.}\ \bibnamefont
  {Pecharsky}}, \bibinfo {author} {\bibfnamefont {K.~A.~Gschneidner}\
  \bibnamefont {Jr}}, \ and\ \bibinfo {author} {\bibfnamefont {D.}~\bibnamefont
  {Fort}},\ }\bibfield  {title} {\enquote {\bibinfo {title} {{Superheating and
  other unusual observations regarding the first order phase transition in
  Dy}},}\ }\href {\doibase 10.1016/1359-6462(96)00225-4} {\bibfield  {journal}
  {\bibinfo  {journal} {Scripta Materialia}\ }\textbf {\bibinfo {volume}
  {35}},\ \bibinfo {pages} {843--848} (\bibinfo {year} {1996})}\BibitemShut
  {NoStop}%
\bibitem [{\citenamefont {Griffel}\ \emph {et~al.}(1954)\citenamefont
  {Griffel}, \citenamefont {Skochdopole},\ and\ \citenamefont
  {Spedding}}]{grif1954a}%
  \BibitemOpen
  \bibfield  {author} {\bibinfo {author} {\bibfnamefont {M.}~\bibnamefont
  {Griffel}}, \bibinfo {author} {\bibfnamefont {R.~E.}\ \bibnamefont
  {Skochdopole}}, \ and\ \bibinfo {author} {\bibfnamefont {F.~H.}\ \bibnamefont
  {Spedding}},\ }\bibfield  {title} {\enquote {\bibinfo {title} {The heat
  capacity of gadolinium from 15 to 355\,k},}\ }\href {\doibase
  10.1103/PhysRev.93.657} {\bibfield  {journal} {\bibinfo  {journal} {Phys.
  Rev.}\ }\textbf {\bibinfo {volume} {93}},\ \bibinfo {pages} {657--661}
  (\bibinfo {year} {1954})}\BibitemShut {NoStop}%
\bibitem [{\citenamefont {Gerstein}\ \emph {et~al.}(1957)\citenamefont
  {Gerstein}, \citenamefont {Griffel}, \citenamefont {Jennings}, \citenamefont
  {Miller}, \citenamefont {Skochdopole},\ and\ \citenamefont
  {Spedding}}]{gers1957a}%
  \BibitemOpen
  \bibfield  {author} {\bibinfo {author} {\bibfnamefont {B.~C.}\ \bibnamefont
  {Gerstein}}, \bibinfo {author} {\bibfnamefont {M.}~\bibnamefont {Griffel}},
  \bibinfo {author} {\bibfnamefont {L.~D.}\ \bibnamefont {Jennings}}, \bibinfo
  {author} {\bibfnamefont {R.~E.}\ \bibnamefont {Miller}}, \bibinfo {author}
  {\bibfnamefont {R.~E.}\ \bibnamefont {Skochdopole}}, \ and\ \bibinfo {author}
  {\bibfnamefont {F.~H.}\ \bibnamefont {Spedding}},\ }\bibfield  {title}
  {\enquote {\bibinfo {title} {Heat capacity of holmium from 15 to 300\,k},}\
  }\href {\doibase 10.1063/1.1743734} {\bibfield  {journal} {\bibinfo
  {journal} {The Journal of Chemical Physics}\ }\textbf {\bibinfo {volume}
  {27}},\ \bibinfo {pages} {394--399} (\bibinfo {year} {1957})}\BibitemShut
  {NoStop}%
\bibitem [{\citenamefont {Boys}\ and\ \citenamefont
  {Legvold}(1968)}]{boys1986a}%
  \BibitemOpen
  \bibfield  {author} {\bibinfo {author} {\bibfnamefont {D.~W.}\ \bibnamefont
  {Boys}}\ and\ \bibinfo {author} {\bibfnamefont {S.}~\bibnamefont {Legvold}},\
  }\bibfield  {title} {\enquote {\bibinfo {title} {Thermal conductivities and
  lorenz functions of dy, er, and lu single crystals},}\ }\href {\doibase
  10.1103/PhysRev.174.377} {\bibfield  {journal} {\bibinfo  {journal} {Phys.
  Rev.}\ }\textbf {\bibinfo {volume} {174}},\ \bibinfo {pages} {377--384}
  (\bibinfo {year} {1968})}\BibitemShut {NoStop}%
\bibitem [{\citenamefont {Hess}\ \emph {et~al.}(2003)\citenamefont {Hess},
  \citenamefont {B\"uchner}, \citenamefont {Ammerahl}, \citenamefont
  {Colonescu}, \citenamefont {Heidrich-Meisner}, \citenamefont {Brenig},\ and\
  \citenamefont {Revcolevschi}}]{hess2003}%
  \BibitemOpen
  \bibfield  {author} {\bibinfo {author} {\bibfnamefont {C.}~\bibnamefont
  {Hess}}, \bibinfo {author} {\bibfnamefont {B.}~\bibnamefont {B\"uchner}},
  \bibinfo {author} {\bibfnamefont {U.}~\bibnamefont {Ammerahl}}, \bibinfo
  {author} {\bibfnamefont {L.}~\bibnamefont {Colonescu}}, \bibinfo {author}
  {\bibfnamefont {F.}~\bibnamefont {Heidrich-Meisner}}, \bibinfo {author}
  {\bibfnamefont {W.}~\bibnamefont {Brenig}}, \ and\ \bibinfo {author}
  {\bibfnamefont {A.}~\bibnamefont {Revcolevschi}},\ }\bibfield  {title}
  {\enquote {\bibinfo {title} {Magnon heat transport in doped
  ${\mathrm{l}\mathrm{a}}_{2}{\mathrm{c}\mathrm{u}\mathrm{o}}_{4}$},}\ }\href
  {\doibase 10.1103/PhysRevLett.90.197002} {\bibfield  {journal} {\bibinfo
  {journal} {Phys. Rev. Lett.}\ }\textbf {\bibinfo {volume} {90}},\ \bibinfo
  {pages} {197002} (\bibinfo {year} {2003})}\BibitemShut {NoStop}%
\bibitem [{\citenamefont {Beaurepaire}\ \emph {et~al.}(1996)\citenamefont
  {Beaurepaire}, \citenamefont {Merle}, \citenamefont {Daunois},\ and\
  \citenamefont {Bigot}}]{beau1996a}%
  \BibitemOpen
  \bibfield  {author} {\bibinfo {author} {\bibfnamefont {E.}~\bibnamefont
  {Beaurepaire}}, \bibinfo {author} {\bibfnamefont {J-C}\ \bibnamefont
  {Merle}}, \bibinfo {author} {\bibfnamefont {A.}~\bibnamefont {Daunois}}, \
  and\ \bibinfo {author} {\bibfnamefont {J-Y}\ \bibnamefont {Bigot}},\
  }\bibfield  {title} {\enquote {\bibinfo {title} {{Ultrafast spin dynamics in
  ferromagnetic nickel}},}\ }\href@noop {} {\bibfield  {journal} {\bibinfo
  {journal} {Physical review letters}\ }\textbf {\bibinfo {volume} {76}},\
  \bibinfo {pages} {4250} (\bibinfo {year} {1996})}\BibitemShut {NoStop}%
\bibitem [{\citenamefont {Biele}\ \emph {et~al.}(2015)\citenamefont {Biele},
  \citenamefont {D'Agosta},\ and\ \citenamefont {Rubio}}]{biel2015}%
  \BibitemOpen
  \bibfield  {author} {\bibinfo {author} {\bibfnamefont {Robert}\ \bibnamefont
  {Biele}}, \bibinfo {author} {\bibfnamefont {Roberto}\ \bibnamefont
  {D'Agosta}}, \ and\ \bibinfo {author} {\bibfnamefont {Angel}\ \bibnamefont
  {Rubio}},\ }\bibfield  {title} {\enquote {\bibinfo {title} {Time-dependent
  thermal transport theory},}\ }\href {\doibase 10.1103/PhysRevLett.115.056801}
  {\bibfield  {journal} {\bibinfo  {journal} {Phys. Rev. Lett.}\ }\textbf
  {\bibinfo {volume} {115}},\ \bibinfo {pages} {056801} (\bibinfo {year}
  {2015})}\BibitemShut {NoStop}%
\bibitem [{\citenamefont {Stamm}\ \emph {et~al.}({2007})\citenamefont {Stamm},
  \citenamefont {Kachel}, \citenamefont {Pontius}, \citenamefont {Mitzner},
  \citenamefont {Quast}, \citenamefont {Holldack}, \citenamefont {Khan},
  \citenamefont {Lupulescu}, \citenamefont {Aziz}, \citenamefont {Wietstruk},
  \citenamefont {Durr},\ and\ \citenamefont {Eberhardt}}]{stam2007a}%
  \BibitemOpen
  \bibfield  {author} {\bibinfo {author} {\bibfnamefont {C.}~\bibnamefont
  {Stamm}}, \bibinfo {author} {\bibfnamefont {T.}~\bibnamefont {Kachel}},
  \bibinfo {author} {\bibfnamefont {N.}~\bibnamefont {Pontius}}, \bibinfo
  {author} {\bibfnamefont {R.}~\bibnamefont {Mitzner}}, \bibinfo {author}
  {\bibfnamefont {T.}~\bibnamefont {Quast}}, \bibinfo {author} {\bibfnamefont
  {K.}~\bibnamefont {Holldack}}, \bibinfo {author} {\bibfnamefont
  {S.}~\bibnamefont {Khan}}, \bibinfo {author} {\bibfnamefont {C.}~\bibnamefont
  {Lupulescu}}, \bibinfo {author} {\bibfnamefont {E.~F.}\ \bibnamefont {Aziz}},
  \bibinfo {author} {\bibfnamefont {M.}~\bibnamefont {Wietstruk}}, \bibinfo
  {author} {\bibfnamefont {H.~A.}\ \bibnamefont {Durr}}, \ and\ \bibinfo
  {author} {\bibfnamefont {W.}~\bibnamefont {Eberhardt}},\ }\bibfield  {title}
  {\enquote {\bibinfo {title} {{Femtosecond modification of electron
  localization and transfer of angular momentum in nickel}},}\ }\href {\doibase
  {10.1038/nmat1985}} {\bibfield  {journal} {\bibinfo  {journal} {{Nature
  Materials}}\ }\textbf {\bibinfo {volume} {{6}}},\ \bibinfo {pages}
  {{740--743}} (\bibinfo {year} {{2007}})}\BibitemShut {NoStop}%
\bibitem [{\citenamefont {Eschenlohr}\ \emph {et~al.}({2013})\citenamefont
  {Eschenlohr}, \citenamefont {Battiato}, \citenamefont {Maldonado},
  \citenamefont {Pontius}, \citenamefont {Kachel}, \citenamefont {Holldack},
  \citenamefont {Mitzner}, \citenamefont {Foehlisch}, \citenamefont
  {Oppeneer},\ and\ \citenamefont {Stamm}}]{esch2013a}%
  \BibitemOpen
  \bibfield  {author} {\bibinfo {author} {\bibfnamefont {A.}~\bibnamefont
  {Eschenlohr}}, \bibinfo {author} {\bibfnamefont {M.}~\bibnamefont
  {Battiato}}, \bibinfo {author} {\bibfnamefont {R.}~\bibnamefont {Maldonado}},
  \bibinfo {author} {\bibfnamefont {N.}~\bibnamefont {Pontius}}, \bibinfo
  {author} {\bibfnamefont {T.}~\bibnamefont {Kachel}}, \bibinfo {author}
  {\bibfnamefont {K.}~\bibnamefont {Holldack}}, \bibinfo {author}
  {\bibfnamefont {R.}~\bibnamefont {Mitzner}}, \bibinfo {author} {\bibfnamefont
  {A.}~\bibnamefont {Foehlisch}}, \bibinfo {author} {\bibfnamefont {P.~M.}\
  \bibnamefont {Oppeneer}}, \ and\ \bibinfo {author} {\bibfnamefont
  {C.}~\bibnamefont {Stamm}},\ }\bibfield  {title} {\enquote {\bibinfo {title}
  {{Ultrafast spin transport as key to femtosecond demagnetization}},}\ }\href
  {\doibase {10.1038/NMAT3546}} {\bibfield  {journal} {\bibinfo  {journal}
  {{Nature Materials}}\ }\textbf {\bibinfo {volume} {{12}}},\ \bibinfo {pages}
  {{332--336}} (\bibinfo {year} {{2013}})}\BibitemShut {NoStop}%
\bibitem [{\citenamefont {Frietsch}\ \emph {et~al.}(2015)\citenamefont
  {Frietsch}, \citenamefont {Bowlan}, \citenamefont {Carley}, \citenamefont
  {Teichmann}, \citenamefont {Wienholdt}, \citenamefont {Hinzke}, \citenamefont
  {Nowak}, \citenamefont {Carva}, \citenamefont {Oppeneer},\ and\ \citenamefont
  {Weinelt}}]{frie2015a}%
  \BibitemOpen
  \bibfield  {author} {\bibinfo {author} {\bibfnamefont {B.}~\bibnamefont
  {Frietsch}}, \bibinfo {author} {\bibfnamefont {J.}~\bibnamefont {Bowlan}},
  \bibinfo {author} {\bibfnamefont {R.}~\bibnamefont {Carley}}, \bibinfo
  {author} {\bibfnamefont {M.}~\bibnamefont {Teichmann}}, \bibinfo {author}
  {\bibfnamefont {S.}~\bibnamefont {Wienholdt}}, \bibinfo {author}
  {\bibfnamefont {D.}~\bibnamefont {Hinzke}}, \bibinfo {author} {\bibfnamefont
  {U.}~\bibnamefont {Nowak}}, \bibinfo {author} {\bibfnamefont
  {K.}~\bibnamefont {Carva}}, \bibinfo {author} {\bibfnamefont {P.~M.}\
  \bibnamefont {Oppeneer}}, \ and\ \bibinfo {author} {\bibfnamefont
  {M.}~\bibnamefont {Weinelt}},\ }\bibfield  {title} {\enquote {\bibinfo
  {title} {{Disparate ultrafast dynamics of itinerant and localized magnetic
  moments in gadolinium metal.}}}\ }\href {\doibase 10.1038/ncomms9262}
  {\bibfield  {journal} {\bibinfo  {journal} {Nature communications}\ }\textbf
  {\bibinfo {volume} {6}},\ \bibinfo {pages} {8262} (\bibinfo {year}
  {2015})}\BibitemShut {NoStop}%
\bibitem [{\citenamefont {Rettig}\ \emph {et~al.}(2015)\citenamefont {Rettig},
  \citenamefont {Dornes}, \citenamefont {Thielemann-Kuehn}, \citenamefont
  {Pontius}, \citenamefont {Zabel}, \citenamefont {Schlagel}, \citenamefont
  {Lograsso}, \citenamefont {Chollet}, \citenamefont {Robert}, \citenamefont
  {Sikorski}, \citenamefont {Song}, \citenamefont {Glownia}, \citenamefont
  {Schuessler-Langeheine}, \citenamefont {Johnson},\ and\ \citenamefont
  {Staub}}]{rett2015a}%
  \BibitemOpen
  \bibfield  {author} {\bibinfo {author} {\bibfnamefont {L.}~\bibnamefont
  {Rettig}}, \bibinfo {author} {\bibfnamefont {C.}~\bibnamefont {Dornes}},
  \bibinfo {author} {\bibfnamefont {N.}~\bibnamefont {Thielemann-Kuehn}},
  \bibinfo {author} {\bibfnamefont {N.}~\bibnamefont {Pontius}}, \bibinfo
  {author} {\bibfnamefont {H.}~\bibnamefont {Zabel}}, \bibinfo {author}
  {\bibfnamefont {D.~L.}\ \bibnamefont {Schlagel}}, \bibinfo {author}
  {\bibfnamefont {T.~A.}\ \bibnamefont {Lograsso}}, \bibinfo {author}
  {\bibfnamefont {M.}~\bibnamefont {Chollet}}, \bibinfo {author} {\bibfnamefont
  {A.}~\bibnamefont {Robert}}, \bibinfo {author} {\bibfnamefont
  {M.}~\bibnamefont {Sikorski}}, \bibinfo {author} {\bibfnamefont
  {S.}~\bibnamefont {Song}}, \bibinfo {author} {\bibfnamefont {J.~M.}\
  \bibnamefont {Glownia}}, \bibinfo {author} {\bibfnamefont {C.}~\bibnamefont
  {Schuessler-Langeheine}}, \bibinfo {author} {\bibfnamefont {S.~L.}\
  \bibnamefont {Johnson}}, \ and\ \bibinfo {author} {\bibfnamefont
  {U.}~\bibnamefont {Staub}},\ }\bibfield  {title} {\enquote {\bibinfo {title}
  {{Itinerant and localized magnetization dynamics in antiferromagnetic Ho}},}\
  }\href {\doibase 10.1103/PhysRevLett.116.257202} {\ \textbf {\bibinfo
  {volume} {257202}},\ \bibinfo {pages} {1--6} (\bibinfo {year} {2015})},\
  \Eprint {http://arxiv.org/abs/1511.05315} {1511.05315} \BibitemShut {NoStop}%
\bibitem [{\citenamefont {Li}\ \emph {et~al.}(2016)\citenamefont {Li},
  \citenamefont {Shelford}, \citenamefont {Shafer}, \citenamefont {Tan},
  \citenamefont {Deng}, \citenamefont {Keatley}, \citenamefont {Hwang},
  \citenamefont {Arenholz}, \citenamefont {van~der Laan}, \citenamefont
  {Hicken},\ and\ \citenamefont {Qiu}}]{li2016a}%
  \BibitemOpen
  \bibfield  {author} {\bibinfo {author} {\bibfnamefont {J.}~\bibnamefont
  {Li}}, \bibinfo {author} {\bibfnamefont {L.~R.}\ \bibnamefont {Shelford}},
  \bibinfo {author} {\bibfnamefont {P.}~\bibnamefont {Shafer}}, \bibinfo
  {author} {\bibfnamefont {A.}~\bibnamefont {Tan}}, \bibinfo {author}
  {\bibfnamefont {J.~X.}\ \bibnamefont {Deng}}, \bibinfo {author}
  {\bibfnamefont {P.~S.}\ \bibnamefont {Keatley}}, \bibinfo {author}
  {\bibfnamefont {C.}~\bibnamefont {Hwang}}, \bibinfo {author} {\bibfnamefont
  {E.}~\bibnamefont {Arenholz}}, \bibinfo {author} {\bibfnamefont
  {G.}~\bibnamefont {van~der Laan}}, \bibinfo {author} {\bibfnamefont {R.~J.}\
  \bibnamefont {Hicken}}, \ and\ \bibinfo {author} {\bibfnamefont {Z.~Q.}\
  \bibnamefont {Qiu}},\ }\bibfield  {title} {\enquote {\bibinfo {title} {Direct
  detection of pure ac spin current by x-ray pump-probe measurements},}\ }\href
  {\doibase 10.1103/PhysRevLett.117.076602} {\bibfield  {journal} {\bibinfo
  {journal} {Phys. Rev. Lett.}\ }\textbf {\bibinfo {volume} {117}},\ \bibinfo
  {pages} {076602} (\bibinfo {year} {2016})}\BibitemShut {NoStop}%
\bibitem [{\citenamefont {Choi}\ \emph {et~al.}()\citenamefont {Choi},
  \citenamefont {Min}, \citenamefont {K.-J.},\ and\ \citenamefont
  {Cahill}}]{choi2014}%
  \BibitemOpen
  \bibfield  {author} {\bibinfo {author} {\bibfnamefont {G.-M.}\ \bibnamefont
  {Choi}}, \bibinfo {author} {\bibfnamefont {B.C.}\ \bibnamefont {Min}},
  \bibinfo {author} {\bibfnamefont {Lee}\ \bibnamefont {K.-J.}}, \ and\
  \bibinfo {author} {\bibfnamefont {D.G.}\ \bibnamefont {Cahill}},\ }\bibfield
  {title} {\enquote {\bibinfo {title} {Spin current generated by thermally
  driven ultrafast demagnetization},}\ }\href@noop {} {\bibinfo  {journal}
  {Nature Communications, volume = {5}, pages = {4334}, year = {2014}, doi =
  {10.1038/ncomms5334},}\ }\BibitemShut {NoStop}%
\bibitem [{\citenamefont {Navirian}\ \emph {et~al.}(2014)\citenamefont
  {Navirian}, \citenamefont {Schick}, \citenamefont {Gaal}, \citenamefont
  {Leitenberger}, \citenamefont {Shayduk},\ and\ \citenamefont
  {Bargheer}}]{navi2013a}%
  \BibitemOpen
\bibfield  {journal} {  }\bibfield  {author} {\bibinfo {author} {\bibfnamefont
  {H.~A.}\ \bibnamefont {Navirian}}, \bibinfo {author} {\bibfnamefont
  {D.}~\bibnamefont {Schick}}, \bibinfo {author} {\bibfnamefont
  {P.}~\bibnamefont {Gaal}}, \bibinfo {author} {\bibfnamefont {W.}~\bibnamefont
  {Leitenberger}}, \bibinfo {author} {\bibfnamefont {R.}~\bibnamefont
  {Shayduk}}, \ and\ \bibinfo {author} {\bibfnamefont {M.}~\bibnamefont
  {Bargheer}},\ }\bibfield  {title} {\enquote {\bibinfo {title} {Thermoelastic
  study of nanolayered structures using time-resolved x-ray diffraction at high
  repetition rate},}\ }\href {\doibase 10.1063/1.4861873} {\bibfield  {journal}
  {\bibinfo  {journal} {Appl. Phys. Lett.}\ }\textbf {\bibinfo {volume}
  {104}},\ \bibinfo {pages} {021906} (\bibinfo {year} {2014})}\BibitemShut
  {NoStop}%
\bibitem [{\citenamefont {Xu}\ \emph {et~al.}(2014)\citenamefont {Xu},
  \citenamefont {Sun}, \citenamefont {Brewe}, \citenamefont {Han},
  \citenamefont {Ho}, \citenamefont {Chen}, \citenamefont {Heald},
  \citenamefont {Zhang},\ and\ \citenamefont {Chow}}]{xu2014}%
  \BibitemOpen
  \bibfield  {author} {\bibinfo {author} {\bibfnamefont {D.~B.}\ \bibnamefont
  {Xu}}, \bibinfo {author} {\bibfnamefont {C.~J.}\ \bibnamefont {Sun}},
  \bibinfo {author} {\bibfnamefont {D.~L.}\ \bibnamefont {Brewe}}, \bibinfo
  {author} {\bibfnamefont {S.-W.}\ \bibnamefont {Han}}, \bibinfo {author}
  {\bibfnamefont {P.}~\bibnamefont {Ho}}, \bibinfo {author} {\bibfnamefont
  {J.~S.}\ \bibnamefont {Chen}}, \bibinfo {author} {\bibfnamefont {S.~M.}\
  \bibnamefont {Heald}}, \bibinfo {author} {\bibfnamefont {X.~Y.}\ \bibnamefont
  {Zhang}}, \ and\ \bibinfo {author} {\bibfnamefont {G.~M.}\ \bibnamefont
  {Chow}},\ }\bibfield  {title} {\enquote {\bibinfo {title} {Spatiotemporally
  separating electron and phonon thermal transport in l10 fept films for heat
  assisted magnetic recording},}\ }\href {\doibase 10.1063/1.4885428}
  {\bibfield  {journal} {\bibinfo  {journal} {Journal of Applied Physics}\
  }\textbf {\bibinfo {volume} {115}},\ \bibinfo {pages} {243907} (\bibinfo
  {year} {2014})},\ \Eprint
  {http://arxiv.org/abs/http://dx.doi.org/10.1063/1.4885428}
  {http://dx.doi.org/10.1063/1.4885428} \BibitemShut {NoStop}%
\bibitem [{\citenamefont {Ho}\ \emph {et~al.}(1974)\citenamefont {Ho},
  \citenamefont {Powell},\ and\ \citenamefont {Liley}}]{ho1974a}%
  \BibitemOpen
  \bibfield  {author} {\bibinfo {author} {\bibfnamefont {Y.~C.}\ \bibnamefont
  {Ho}}, \bibinfo {author} {\bibfnamefont {R.~W.}\ \bibnamefont {Powell}}, \
  and\ \bibinfo {author} {\bibfnamefont {P.~E}\ \bibnamefont {Liley}},\
  }\bibfield  {title} {\enquote {\bibinfo {title} {{Thermal Conductivity of the
  Elements: A Comprehensive Review}},}\ }\bibfield  {booktitle} {\emph
  {\bibinfo {booktitle} {Journal of Physical and Chemical Reference Data}},\
  }\href@noop {} {\ \textbf {\bibinfo {volume} {3}},\ \bibinfo {pages} {810}
  (\bibinfo {year} {1974})}\BibitemShut {NoStop}%
\bibitem [{\citenamefont {Dobrovinskaya}\ \emph {et~al.}(2009)\citenamefont
  {Dobrovinskaya}, \citenamefont {Lytvynov},\ and\ \citenamefont
  {Pishchik}}]{dobr2009a}%
  \BibitemOpen
  \bibfield  {author} {\bibinfo {author} {\bibfnamefont {E.~R.}\ \bibnamefont
  {Dobrovinskaya}}, \bibinfo {author} {\bibfnamefont {L.~A.}\ \bibnamefont
  {Lytvynov}}, \ and\ \bibinfo {author} {\bibfnamefont {V.}~\bibnamefont
  {Pishchik}},\ }\href@noop {} {\emph {\bibinfo {title} {"Sapphire", Micro- and
  Opto-Electronic Materials, Structures, and Systems}}}\ (\bibinfo  {publisher}
  {Springer},\ \bibinfo {year} {2009})\BibitemShut {NoStop}%
\bibitem [{\citenamefont {Dumesnil}\ \emph {et~al.}(1995)\citenamefont
  {Dumesnil}, \citenamefont {Dufour}, \citenamefont {Mangin}, \citenamefont
  {Marchal},\ and\ \citenamefont {Hennion}}]{dume1995a}%
  \BibitemOpen
  \bibfield  {author} {\bibinfo {author} {\bibfnamefont {K.}~\bibnamefont
  {Dumesnil}}, \bibinfo {author} {\bibfnamefont {C.}~\bibnamefont {Dufour}},
  \bibinfo {author} {\bibfnamefont {Ph}~\bibnamefont {Mangin}}, \bibinfo
  {author} {\bibfnamefont {G.}~\bibnamefont {Marchal}}, \ and\ \bibinfo
  {author} {\bibfnamefont {M.}~\bibnamefont {Hennion}},\ }\bibfield  {title}
  {\enquote {\bibinfo {title} {{Magnetoelastic and Exchange Contributions to
  the Helical-Ferromagnetic Transition in Dysprosium Epitaxial Films}},}\
  }\href@noop {} {\bibfield  {journal} {\bibinfo  {journal} {EPL (Europhysics
  Letters)}\ }\textbf {\bibinfo {volume} {31}},\ \bibinfo {pages} {43}
  (\bibinfo {year} {1995})}\BibitemShut {NoStop}%
\bibitem [{\citenamefont {Navirian}\ \emph {et~al.}(2012)\citenamefont
  {Navirian}, \citenamefont {Shayduk}, \citenamefont {Leitenberger},
  \citenamefont {Goldshteyn}, \citenamefont {Gaal},\ and\ \citenamefont
  {Bargheer}}]{navi2012a}%
  \BibitemOpen
  \bibfield  {author} {\bibinfo {author} {\bibfnamefont {H.}~\bibnamefont
  {Navirian}}, \bibinfo {author} {\bibfnamefont {R.}~\bibnamefont {Shayduk}},
  \bibinfo {author} {\bibfnamefont {W.}~\bibnamefont {Leitenberger}}, \bibinfo
  {author} {\bibfnamefont {J.}~\bibnamefont {Goldshteyn}}, \bibinfo {author}
  {\bibfnamefont {P.}~\bibnamefont {Gaal}}, \ and\ \bibinfo {author}
  {\bibfnamefont {M.}~\bibnamefont {Bargheer}},\ }\bibfield  {title} {\enquote
  {\bibinfo {title} {Synchrotron-based ultrafast x-ray diffraction at high
  repetition rates},}\ }\href {\doibase 10.1063/1.4727872} {\bibfield
  {journal} {\bibinfo  {journal} {Rev. Sci. Instrum.}\ }\textbf {\bibinfo
  {volume} {83}},\ \bibinfo {pages} {063303} (\bibinfo {year}
  {2012})}\BibitemShut {NoStop}%
\bibitem [{\citenamefont {Reinhardt}\ \emph {et~al.}(2016)\citenamefont
  {Reinhardt}, \citenamefont {Koc}, \citenamefont {Leitenberger}, \citenamefont
  {Gaal},\ and\ \citenamefont {Bargheer}}]{rein2016a}%
  \BibitemOpen
  \bibfield  {author} {\bibinfo {author} {\bibfnamefont {M.}~\bibnamefont
  {Reinhardt}}, \bibinfo {author} {\bibfnamefont {A.}~\bibnamefont {Koc}},
  \bibinfo {author} {\bibfnamefont {W.}~\bibnamefont {Leitenberger}}, \bibinfo
  {author} {\bibfnamefont {P.}~\bibnamefont {Gaal}}, \ and\ \bibinfo {author}
  {\bibfnamefont {M.}~\bibnamefont {Bargheer}},\ }\bibfield  {title} {\enquote
  {\bibinfo {title} {{Optimized spatial overlap in optical pump{--}X-ray probe
  experiments with high repetition rate using laser-induced surface
  distortions}},}\ }\href {\doibase 10.1107/s1600577515024443} {\bibfield
  {journal} {\bibinfo  {journal} {Journal of Synchrotron Radiation}\ }\textbf
  {\bibinfo {volume} {23}},\ \bibinfo {pages} {474--479} (\bibinfo {year}
  {2016})}\BibitemShut {NoStop}%
\bibitem [{\citenamefont {Schick}\ \emph {et~al.}(2013)\citenamefont {Schick},
  \citenamefont {Shayduk}, \citenamefont {Bojahr}, \citenamefont {Herzog},
  \citenamefont {von Korff~Schmising}, \citenamefont {Gaal},\ and\
  \citenamefont {Bargheer}}]{schi2013a}%
  \BibitemOpen
  \bibfield  {author} {\bibinfo {author} {\bibfnamefont {D.}~\bibnamefont
  {Schick}}, \bibinfo {author} {\bibfnamefont {R.}~\bibnamefont {Shayduk}},
  \bibinfo {author} {\bibfnamefont {A.}~\bibnamefont {Bojahr}}, \bibinfo
  {author} {\bibfnamefont {M.}~\bibnamefont {Herzog}}, \bibinfo {author}
  {\bibfnamefont {C.}~\bibnamefont {von Korff~Schmising}}, \bibinfo {author}
  {\bibfnamefont {P.}~\bibnamefont {Gaal}}, \ and\ \bibinfo {author}
  {\bibfnamefont {M.}~\bibnamefont {Bargheer}},\ }\bibfield  {title} {\enquote
  {\bibinfo {title} {Ultrafast reciprocal-space mapping with a convergent
  beam},}\ }\href {\doibase 10.1107/s0021889813020013} {\bibfield  {journal}
  {\bibinfo  {journal} {Journal of Applied Crystallography}\ }\textbf {\bibinfo
  {volume} {46}},\ \bibinfo {pages} {1372--1377} (\bibinfo {year}
  {2013})}\BibitemShut {NoStop}%
\bibitem [{\citenamefont {von Reppert}\ \emph {et~al.}(2016)\citenamefont {von
  Reppert}, \citenamefont {Pudell}, \citenamefont {Koc}, \citenamefont
  {Reinhardt}, \citenamefont {Leitenberger}, \citenamefont {Dumesnil},
  \citenamefont {Zamponi},\ and\ \citenamefont {Bargheer}}]{repp2016a}%
  \BibitemOpen
  \bibfield  {author} {\bibinfo {author} {\bibfnamefont {A.}~\bibnamefont {von
  Reppert}}, \bibinfo {author} {\bibfnamefont {J.}~\bibnamefont {Pudell}},
  \bibinfo {author} {\bibfnamefont {A.}~\bibnamefont {Koc}}, \bibinfo {author}
  {\bibfnamefont {M.}~\bibnamefont {Reinhardt}}, \bibinfo {author}
  {\bibfnamefont {W.}~\bibnamefont {Leitenberger}}, \bibinfo {author}
  {\bibfnamefont {K.}~\bibnamefont {Dumesnil}}, \bibinfo {author}
  {\bibfnamefont {F.}~\bibnamefont {Zamponi}}, \ and\ \bibinfo {author}
  {\bibfnamefont {M.}~\bibnamefont {Bargheer}},\ }\bibfield  {title} {\enquote
  {\bibinfo {title} {Persistent nonequilibrium dynamics of the thermal energies
  in the spin and phonon systems of an antiferromagnet},}\ }\href@noop {}
  {\bibfield  {journal} {\bibinfo  {journal} {Structural Dynamics}\ }\textbf
  {\bibinfo {volume} {3}},\ \bibinfo {eid} {054302} (\bibinfo {year}
  {2016})}\BibitemShut {NoStop}%
\bibitem [{\citenamefont {Smith}\ and\ \citenamefont
  {Gjevre}(1960)}]{smit1960a}%
  \BibitemOpen
  \bibfield  {author} {\bibinfo {author} {\bibfnamefont {J.~F.}\ \bibnamefont
  {Smith}}\ and\ \bibinfo {author} {\bibfnamefont {J.~A.}\ \bibnamefont
  {Gjevre}},\ }\bibfield  {title} {\enquote {\bibinfo {title} {{Elastic
  constants of yttrium single crystals in the temperature range 4.2-400\,K}},}\
  }\href {\doibase 10.1063/1.1735657} {\bibfield  {journal} {\bibinfo
  {journal} {Journal of Applied Physics}\ }\textbf {\bibinfo {volume} {31}},\
  \bibinfo {pages} {645--647} (\bibinfo {year} {1960})}\BibitemShut {NoStop}%
\bibitem [{\citenamefont {Rosen}\ and\ \citenamefont
  {Klimker}(1970)}]{rose1970a}%
  \BibitemOpen
  \bibfield  {author} {\bibinfo {author} {\bibfnamefont {M.}~\bibnamefont
  {Rosen}}\ and\ \bibinfo {author} {\bibfnamefont {H.}~\bibnamefont
  {Klimker}},\ }\bibfield  {title} {\enquote {\bibinfo {title}
  {{Low-temperature elasticity and magneto-elasticity of dysprosium single
  crystals}},}\ }\href {\doibase 10.1103/PhysRevB.1.3748} {\bibfield  {journal}
  {\bibinfo  {journal} {Physical Review B}\ }\textbf {\bibinfo {volume} {1}},\
  \bibinfo {pages} {3748--3756} (\bibinfo {year} {1970})}\BibitemShut {NoStop}%
\bibitem [{\citenamefont {{Keith J. Carroll}}(1965)}]{caro1965a}%
  \BibitemOpen
  \bibfield  {author} {\bibinfo {author} {\bibnamefont {{Keith J. Carroll}}},\
  }\bibfield  {title} {\enquote {\bibinfo {title} {{Elastic Constants of
  Niobium from 4.2 to 300\,K}},}\ }\href {\doibase 10.1063/1.1703072}
  {\bibfield  {journal} {\bibinfo  {journal} {Journal of Applied Physics}\
  }\textbf {\bibinfo {volume} {36}},\ \bibinfo {pages} {3689} (\bibinfo {year}
  {1965})}\BibitemShut {NoStop}%
\bibitem [{\citenamefont {Jennings}\ \emph {et~al.}(1960)\citenamefont
  {Jennings}, \citenamefont {Miller},\ and\ \citenamefont
  {Spedding}}]{jenn1960a}%
  \BibitemOpen
  \bibfield  {author} {\bibinfo {author} {\bibfnamefont {L.~D.}\ \bibnamefont
  {Jennings}}, \bibinfo {author} {\bibfnamefont {R.~E.}\ \bibnamefont
  {Miller}}, \ and\ \bibinfo {author} {\bibfnamefont {F.~H.}\ \bibnamefont
  {Spedding}},\ }\bibfield  {title} {\enquote {\bibinfo {title} {Lattice heat
  capacity of the rare earths. heat capacities of yttrium and lutetium from
  15–350\,k},}\ }\href {\doibase 10.1063/1.1731516} {\bibfield  {journal}
  {\bibinfo  {journal} {The Journal of Chemical Physics}\ }\textbf {\bibinfo
  {volume} {33}},\ \bibinfo {pages} {1849--1852} (\bibinfo {year}
  {1960})}\BibitemShut {NoStop}%
\bibitem [{\citenamefont {von Reppert}(2015)}]{repp2015a}%
  \BibitemOpen
  \bibfield  {author} {\bibinfo {author} {\bibfnamefont {Alexander}\
  \bibnamefont {von Reppert}},\ }\emph {\bibinfo {title} {{Ultrafast
  magnetostriction in dysprosium studied by femtosecond X-Ray diffraction}}},\
  \href
  {http://aigaionembed.udkm.physik.uni-potsdam.de/attachments/single/1382}
  {Master's thesis},\ \bibinfo  {school} {Universit{\"a}t Potsdam} (\bibinfo
  {year} {2015})\BibitemShut {NoStop}%
\bibitem [{\citenamefont {Schick}\ \emph {et~al.}(2014)\citenamefont {Schick},
  \citenamefont {Bojahr}, \citenamefont {Herzog}, \citenamefont {Shayduk},
  \citenamefont {von Korff~Schmising},\ and\ \citenamefont
  {Bargheer}}]{schi2014c}%
  \BibitemOpen
  \bibfield  {author} {\bibinfo {author} {\bibfnamefont {D.}~\bibnamefont
  {Schick}}, \bibinfo {author} {\bibfnamefont {A.}~\bibnamefont {Bojahr}},
  \bibinfo {author} {\bibfnamefont {M.}~\bibnamefont {Herzog}}, \bibinfo
  {author} {\bibfnamefont {R.}~\bibnamefont {Shayduk}}, \bibinfo {author}
  {\bibfnamefont {C.}~\bibnamefont {von Korff~Schmising}}, \ and\ \bibinfo
  {author} {\bibfnamefont {M.}~\bibnamefont {Bargheer}},\ }\bibfield  {title}
  {\enquote {\bibinfo {title} {{UDKM1DSIM---A simulation toolkit for 1D
  ultrafast dynamics in condensed matter}},}\ }\href@noop {} {\bibfield
  {journal} {\bibinfo  {journal} {Computer Physics Communications}\ }\textbf
  {\bibinfo {volume} {185}},\ \bibinfo {pages} {651--660} (\bibinfo {year}
  {2014})}\BibitemShut {NoStop}%
\end{thebibliography}
%
\end{document}